\documentclass[12pt]{article}
\usepackage{a4}
\usepackage{bbm}

\usepackage[english]{babel}
\usepackage{amsfonts}
\usepackage{pslatex}
\usepackage[latin1]{inputenc}
\usepackage[T1]{fontenc}

\ifx\pdfoutput\undefined
\usepackage{hyperref}
\usepackage{graphicx}
\else
\usepackage[pdftex]{hyperref}
\usepackage[pdftex]{graphicx}
\fi

\makeatletter
\def\@sect#1#2#3#4#5#6[#7]#8{\ifnum #2>\c@secnumdepth
  \def\@svsec{}\else 
  \refstepcounter{#1}\edef\@svsec{\csname the#1\endcsname.\hskip0.5em}\fi
  \@tempskipa #5\relax
  \ifdim \@tempskipa>\z@
    \begingroup 
      #6\relax
      \@hangfrom{\hskip #3\relax\@svsec}{\interlinepenalty \@M #8\par}%
    \endgroup
    \csname #1mark\endcsname{#7}\addcontentsline
      {toc}{#1}{\ifnum #2>\c@secnumdepth \else
        \protect\numberline{\csname the#1\endcsname}\fi #7}%
  \else
    \def\@svsechd{#6\hskip #3\@svsec #8\csname #1mark\endcsname
      {#7}\addcontentsline{toc}{#1}{\ifnum #2>\c@secnumdepth \else
        \protect\numberline{\csname the#1\endcsname}\fi #7}}%
  \fi \@xsect{#5}}
\@addtoreset{equation}{section}
\@addtoreset{figure}{section}
\makeatother

\renewcommand\theequation{\ifnum \value{section}>0
 \arabic{section}.\arabic{equation}%
\else
\arabic{equation}%
\fi}
\renewcommand\thefigure{\ifnum \value{section}>0
 \arabic{section}.\arabic{figure}%
\else
\arabic{figure}%
\fi}

\def\one{\mathbbm{1}}
\def\tr{{\rm Tr}}
\def\cO{{\cal O}}
\def\Eq#1{Eq.~(\ref{eq:#1})}
\def\e{\varepsilon}

\begin{document}
\thispagestyle{empty}
\hfill  CERN-PH-TH/2004-246
\vspace*{3cm}
\begin{center}
{\Large\bf Maximizing the spin correlation of top quark pairs produced at
  the Large Hadron Collider} 
\vspace*{0.5cm}

{\large Peter Uwer}\\[0.3cm]
CERN, Department of Physics, Theory Division,\\
  CH-1211 Geneva 23, 
    Switzerland
\end{center}
\vspace*{1.5cm}
\begin{center}
{\bf Abstract:}\\[0.3cm]
\parbox{0.8\textwidth}{
The measurement of top quark spin correlation is an important tool 
for precise studies of top quark interactions. 
In this letter I construct a quantization
axis maximizing the spin correlation at the LHC within the Standard
Model. Using this axis a spin correlation of 48\% or even more, on applying
additional cuts, can be reached. This represents a significant
improvement compared to the helicity bases studied thus far.}
\end{center}

\newpage
\setcounter{page}{1}
\section{Introduction}
At the Large Hadron Collider (LHC) at CERN a huge number of top quark
pairs will be produced. In the low luminosity run, production
of around 8 million top quark pairs per year can be anticipated. 
This large number
of top quarks allows very precise measurements in the top sector. 
In particular, we can verify to high accuracy that the top quark has
indeed the quantum numbers predicted by the Standard
Model. Furthermore, given the high energy scale involved in top quark 
reactions, top quark physics is also an ideal laboratory to search for
new physics. For example, we may search for new $s$-channel 
resonances which may couple strongly to the top quark. To study the 
properties of such a hypothetical resonance the top quark spin
correlation is a suitable tool. In particular, this may help to
disentangle the nature of the intermediate resonance. 
It is important to keep in mind that the top quark is unique among the 
quarks because it decays before it can hadronize \cite{Bigi:1986jk}. 
The spin information
is thus not diluted by hadronization. In the Standard Model where the 
top decays predominantly via the parity violating weak interaction, the spin
information is transferred to the angular distribution of the decay
products. 
The top polarization is thus a `good observable' in the sense that it is
experimentally accessible through a detailed study of the decay
products \cite{Kuhn:1983ix}. 
The aforementioned  spin correlation of top quark pairs can be
defined by \cite{Stelzer:1995gc}
\begin{equation}
  C = {\sigma_{t\bar t}(\uparrow\uparrow)
    +\sigma_{t\bar t}(\downarrow\downarrow) 
    -\sigma_{t\bar t}(\uparrow\downarrow)
    -\sigma_{t\bar t}(\uparrow\downarrow)\over 
    \sigma_{t\bar t}(\uparrow\uparrow)
    +\sigma_{t\bar t}(\downarrow\downarrow) 
    +\sigma_{t\bar t}(\uparrow\downarrow)
    +\sigma_{t\bar t}(\uparrow\downarrow)},
  \label{eq:SpinCorrelation}
\end{equation}
where $\sigma_{t\bar t}(\uparrow\!\!\!\!/\!\!\!\!\downarrow\,\,
\uparrow\!\!\!\!/\!\!\!\!\downarrow)$ denotes the cross section for
the production of a top quark pair ($+X$, beyond leading-order) 
with spins up or down with respect to a specific quantization axis.
In fact, given that already in the 
Standard Model the spins of the top and antitop are correlated, the
spin correlation is also an interesting observable to test the
details of Standard Model top quark interactions with high accuracy.
The main production processes in the Standard Model for top quark 
pair production in hadronic collisions are the quark-antiquark
annihilation process and the gluon fusion process. While the first
dominates top quark pair production at the Tevatron, the latter
dominates top quark pair production at the LHC.
For top quark pairs produced in quark-antiquark annihilation it is
well known that an optimal quantization axis exists --- the so-called
`off-diagonal' axis --- for which  the top spins are 100\%
correlated \cite{Mahlon:1997uc}. 
Given that at the Tevatron roughly 80\% of the top quark
pairs are produced in quark-antiquark annihilation it is thus
sufficient to chose this axis to obtain a large value for  
the spin correlation.
For the gluon process --- as we will see later --- no such optimal
quantization axis exists. Although no optimal axis exists it is still useful to
find  an axis for which the correlation is at least `maximal'. Such an
axis might be used to improve the significance with which the spin
correlation can be established at the LHC. In this
letter I  describe the construction of such an axis in detail.
Note that in the following I will restrict myself to the top quark
final state. Details on how to measure the spin correlation at
the level of the observable decay products can be found for example in 
\cite{Barger:1988jj,Arens:1992fg,Arens:1992wh,Stelzer:1995gc,Brandenburg:1996df}. 

\section{Maximizing the spin correlation}
In this section I discuss how to maximize the top quark spin
correlation in the Standard Model.
To study spin effects in quantum mechanics a convenient tool is 
the spin density matrix at the parton level \cite{Bernreuther:1993hq}.
The most general form of the spin density matrix $\rho$ for top quark pair
production is given by 
\begin{equation}
  \rho = A\, \one\otimes\one
       + B_i^t \, \sigma_i\otimes \one
       + B_i^{\bar t}\, \one \otimes \sigma_i
       + C_{ij}\, \sigma_i\otimes \sigma_j
       \label{eq:rho}
\end{equation}
where $\sigma_i$ are the Pauli matrices.
The operator ${\sigma_i\over 2}\otimes \one$ 
($\one \otimes {\sigma_i\over 2}$) denotes the
spin operator of the top (anti-)quark in its rest frame. 
The
different contributions to the spin density matrix have a very simple
interpretation. The first contribution is essentially nothing 
but the differential cross section for top quark
pair production at the parton level:
\begin{equation}
  {d \sigma_{t \bar t}\over d\cos(\vartheta)} = {\beta \over 8\pi s} A ,
\end{equation}
where $s$ describes the partonic center of mass energy and 
$\beta$ is the velocity of the top quark in the partonic center of
mass system. The scattering angle of the top quark with respect to 
the beam is given by $\vartheta$.
The second and third terms in \Eq{rho} describe the polarization of the top and
antitop quark. The last term may parameterize a correlation between the 
spins of the top and the antitop. Note that a non-vanishing $C_{ij}$ does
not necessarily mean that the spins are correlated. Only in the
absence of polarization does a non-vanishing $C_{ij}$ directly signify 
spin correlation. 
The spin density matrix as given in \Eq{rho} above is not normalized. 
This has to be
taken into account when calculating the expectation values of 
spin observables:
\begin{equation}
  \langle \cO \rangle_\rho = {\tr[\cO \rho]\over \tr[\rho]} = 
  {\tr[\cO \rho]\over \int\! d{\rm Lips}\,\, 4A}. 
\end{equation}
Note that taking the trace also includes a phase space integration
$\int\! d{\rm Lips}$ over the lorentz-invariant phase space.
If the interaction responsible for the production of the top quark
pairs satisfies additional symmetries, the explicit form of the spin
density matrix can be further constrained \cite{Bernreuther:1993hq}. 
For top quark pair
production at a hadron collider, where the responsible interaction is
QCD, we can immediately conclude that at leading-order no polarization
is allowed ($B_i^t=B_i^{\bar t}=0$) due to the parity invariance of
QCD. At the one-loop level, a tiny polarization transverse to the
scattering plane is induced by absorptive parts
\cite{Bernreuther:1995cx}. 
Also the explicit
form of the matrix $C$ is constrained by the symmetries of QCD 
\cite{Bernreuther:1993hq}:
\begin{equation}
  C_{ij} = c_0 \delta_{ij} + \hat p_i \hat p_j c_{4} 
  + \hat k_i \hat k_j c_{5}
  + (\hat k_i \hat p_j + \hat p_i \hat k_j)  c_{6}.
\end{equation}
Here $\bf \hat p$ is the direction of the incoming beam and $\bf\hat k$ is the
direction of the outgoing top quark.
Other structures one could think of, for example
\begin{equation}
  \e_{ijk} ( c_1 \hat p_k + c_2 \hat k_k + c_3 \hat n_k),
\end{equation}
where ${\bf \hat n}$ is given by
\begin{equation}
  {\bf \hat n} = {{\bf \hat p}\times {\bf \hat k}\over 
    |{\bf \hat p}\times {\bf \hat k}|},
\end{equation}
are forbidden in QCD due to discrete symmetries. In leading-order QCD
the spin density matrix is thus completely determined by the 
functions $A, c_0, c_4, c_5,c_6$.
For quark-antiquark annihilation they are given by \cite{Bernreuther:1993hq}:
\def\betaq{{\beta^2}}
 \begin{eqnarray}
    A^q&=&\kappa_q\,\*(2-(1-z^2)\*\beta^2),\\
    c^q_{0}&=&-\kappa_q\*(1-z^2)\*\beta^2,\\
    c^q_{4}&=&2\*\kappa_q,\\
    c^q_{5}&=&2\*\kappa_q\*\bigg((1-z^2)\*\beta^2
    +2\*z^2\*\bigg[1-\sqrt{1-\beta^2}\bigg]
    \bigg),\\
    c^q_{6}&=&-2\*\kappa_q\*z \*\bigg(1-\sqrt{1-\beta^2}\bigg),
  \end{eqnarray}
with 
\begin{equation}
  \kappa_q = \pi^2\*\alpha_s^2\,\*\frac{N^2-1}{N^2}\stackrel{N=3}{=} 
  {8\over 9}\*\pi^2\*\alpha_s^2,
\end{equation}
where $\alpha_s$ is the QCD coupling constant and
$N$ denotes the number of colours.
For the gluon fusion process the functions $A, c_0, c_4, c_5,c_6$ 
are given by \cite{Bernreuther:1993hq}:
 \begin{eqnarray}
      A^g&=&2\*\kappa_g\*\left(1+2\*\beta^2\*(1-\beta^2)\*(1-z^2)
        -z^4\*\beta^4\right),\\
      c^g_{0}&=&-2\*\kappa_g\*\left(1-2\*\beta^2+2\*(1-z^2)\*\beta^4
      +z^4\*\beta^4\right),\\
      c^g_{4}&=&4\*\kappa_g\*(1-z^2)\*\beta^2,\\
      c^g_{5}&=&4\*\kappa_g\*\ \betaq
      \*\bigg(-2\*z^2\*(1-z^2)\*\sqrt{1-\betaq}
      +2\*(z^2+\betaq)\*(1-z^2)-1+\betaq\*z^4\bigg),\\
      c^g_{6}&=&-4\*\kappa_g\* z\* (1-z^2)\*\beta^2\* 
      \bigg(1-\sqrt{1-\beta^2}\bigg),
  \end{eqnarray}
with
\begin{equation}
    \kappa_g = \frac{\pi^2\alpha_s^2}{(1-z^2\beta^2)^2}
    \frac{(N^2-2+N^2z^2\beta^2)}{N(N^2-1)}\stackrel{N=3}{=}
    {1\over 24}\*\pi^2\*\alpha_s^2\*{7+9z^2\beta^2 \over 
      (1-z^2\*\beta^2)^2}.
  \end{equation}
In the absence of polarization, the spin correlation at the parton
level as defined in 
\Eq{SpinCorrelation} is just given by
\begin{equation}
  C =  4 \langle ({\bf a}\cdot {\bf s}_t)({\bf b}\cdot {\bf s}_{\bar t})\rangle
  = {\int\! {\rm Lips} \,\,a_i C_{ij} b_j \over \int\! {\rm Lips}\,\, A},
\end{equation}
where the quantization axis of the (anti)top quark is described by the 
normalized vector $\bf a$ ($\bf b$). 
It is now clear how one can maximize the spin correlation: just
determine the maximal eigenvalue of the matrix $C$ and choose $\bf a$ and
$\bf b$ equal to the corresponding eigenvector, including an additional
sign if the eigenvalue is negative. Note that the matrix $C$ is
symmetric so that this procedure can always be carried out.
Without loss of generality we may choose for the moment a coordinate
frame in which $\bf \hat p$ and $\bf \hat k$ are given by:
\begin{equation}
  {\bf \hat p}=\left(
    \begin{array}[htbp]{c}
      0\\
      0\\
      1\\
    \end{array}
\right)
\quad\mbox{and}\quad
{\bf \hat k}=\left(
    \begin{array}[htbp]{c}
      \sqrt{1-z^2}\\
      0\\
      z\\
    \end{array}
\right),
\end{equation}
with $z=\cos(\vartheta)$. Using this specific coordinate frame the
matrix $C$ reads:
\begin{equation}
  C = \left(
    \begin{array}[htbp]{ccc}
       c_0 + (1-z^2)c_5& 0& z\sqrt{1-z^2}c_5 + \sqrt{1-z^2} c_6\\
      0 & c_0 & 0 \\
      z\sqrt{1-z^2}c_5 +\sqrt{1-z^2} c_6 & 0 & c_0 + c_4 + z^2 c_5 + 2
      z c_6
    \end{array}
  \right).
\end{equation}
It is straightforward to determine the eigenvalues:
\begin{equation}
  c_0, c_0 + {1\over 2}\* c_4 + {1\over 2}\* c_5 + z\* c_6
  \pm {1\over 2}\*\sqrt{c_5^2+c_4^2+4\*c_6^2 
    -2\*c_4\*c_5 +4\*z\*c_5\*c_6+4\*z\*c_4\*c_6
    +4\*z^2\*c_5\*c_4}.
\end{equation}
The corresponding eigenvectors are given by
\begin{equation}
  {\bf e}_1= {\bf \hat p}\times {\bf \hat k},
\end{equation}
and
\begin{eqnarray}
  {\bf e}_{\pm} &=&\left({1\over 2}\*c_4-{1\over 2}\*c_5 
    \pm {1\over 2}\* \sqrt{c_5^2+c_4^2+4\*c_6^2
      -2\*c_4\*c_5 +4\*z\*c_5\*c_6+4\*z\*c_6\*c_4
      +4\*z^2\*c_5\*c_4}\right) {\bf \hat p}\nonumber\\
  &+& (z\*c_5+c_6) {\bf \hat k}
\end{eqnarray}
Note that the eigenvectors are not normalized to one.
The only thing that remains to be done is to determine which of the
eigenvalues is the largest. Clearly this will depend on the
initial state. For the quark-antiquark annihilation process the
largest eigenvalue in the entire kinematical region is given by
the one where the square root enters with a plus sign. 
Using the explicit form for $c_0^q$--$c_6^q$ I find
\begin{eqnarray}
  \lambda_{\rm max}^q &=& 
   c^q_0 + {1\over 2}\* c^q_4 + {1\over 2}\* c^q_5 + z\* c^q_6\nonumber\\
  &+& {1\over 2}\*\sqrt{{c^q_5}^2+{c^q_4}^2+4\*{c^q_6}^2
    +4\*z\*c^q_5\*c^q_6
    -2\*c^q_4\*c^q_5+4\*z\*c^q_4\*c^q_6
    +4\*z^2\*c^q_5\*c^q_4}\nonumber \\
  &=& A^q.
\end{eqnarray}
I thus reproduce the well known result that the maximal axis yield
100\% spin correlation for the quark-antiquark annihilation 
sub-process \cite{Mahlon:1997uc}.
The corresponding eigenvector is than given by
\begin{equation}
  {\bf e}^q_+ \sim  {{\bf p} + (\gamma-1) z {\bf \hat k}\over 
    \sqrt{1+ z^2\*(\gamma^2-1)}}
\end{equation}
in agreement with ref.~\cite{Mahlon:1997uc}. 
Given that at the Tevatron most of the top
quark pairs are produced in quark-antiquark annihilation this axis
will produce an almost optimal value for the spin correlation. 
At the LHC, as mentioned earlier, gluon fusion is
the dominant process.
Unfortunately, for the gluon channel no such compact expression for
the axis maximizing the spin correlation exists. Nevertheless, the
axis can be constructed on an event by event basis. To do so one first 
calculates the eigenvalues of the $C$ matrix for the gluon fusion
process for the event. 
One then determines which one has the largest absolute value.
The quantization axis is then given by the corresponding eigenvector to that
eigenvalue. If the eigenvalue is negative one introduces an additional
sign in the quantization axis of the top or the antitop quark. The quantization
bases constructed in this way will yield an `optimal' value
for the spin correlation at the LHC. By explicitly calculating the
eigenvalues in terms of $z$ and $\beta$ one can also show that none of them
is equal to $A_g$. This implies that for the gluon fusion process no optimal
axis for which the spins are 100\% correlated exists.

\section{Numerical results}
In this section I present results for the spin correlation at the LHC
using the maximal axis derived above. At the LHC it is known that
QCD corrections do not significantly change the spin correlation 
\cite{Bernreuther:2004jv},
therefore I will only discuss leading-order predictions in what
follows. As input I use $m_t=178$ GeV. Note that the spin correlation
only depends on the QCD coupling constant $\alpha_s$ through the 
parton distribution functions. There is no explicit dependence on
$\alpha_s$. For the parton distribution functions 
I use CTEQ6.1L \cite{Stump:2003yu}.
The factorization scale $\mu_f$ is set to $\mu_f=m_t$. 
\begin{table}[htbp]
  \begin{center}
    \leavevmode
    \begin{tabular}[htbp]{c|c|c|c}
      $C_{\rm hel}$  & $C_{\rm max}$ & $C_{\rm hel}$ with cut  
      & $C_{\rm max}$ with cut\\ \hline
      0.318 & 0.484 &0.453 &0.502 \\ \hline
    \end{tabular}
    \caption{Top quark spin correlation at the LHC using different quantization axes. }
    \label{tab:Results}
  \end{center}
\end{table}
The results are shown in Table~\ref{tab:Results}. Using the proposed
axis a spin correlation of almost 50\% is obtained. It is known \cite{Mahlon:1995zn,Pralavorio} that
applying an additional cut on the $t\bar t$ invariant mass can improve  
the observed spin correlation significantly. 
In the two last columns the influence of a  cut, 
\begin{equation}
  (k_t + k_{\bar t})^ 2 < 550\mbox{ GeV},
\end{equation}
is studied. A further increase of the correlation is observed,
although in that case it might be easier to use the helicity bases.
Given that the spin correlation is defined as a ratio of two cross
sections it can be expected that the factorization scale dependence
cancels to a large extent. Indeed varying the factorization scale from
$\mu_f = m_t/10$ to $\mu_f = 10 m_t$ the value for $C_{\rm max}$
changes only from 50.2\% to 46.6\%.  The scale dependence could be
reduced further by including the next-to-leading order corrections 
\cite{Bernreuther:2004jv}.

\section{Conclusion}
In this letter I constructed a quantization axis for which
the spin correlation of top quark pairs produced by gluon fusion is
maximal. Given that around 90\% of the top quark pairs at the LHC are
produced via gluon fusion, this axis will yield an almost maximal
value at the LHC. In leading-order using the CTEQ6.1L the proposed
axis yields a spin correlation of 48\%. 
An additional cut on the $t\bar t$ invariant mass
can be used to further increase the correlation. Using 
$(k_t+k_{\bar t})^2 < 550$ GeV increases the spin correlation $C_{\rm
  max}$ by 2 \%. The use of the proposed axis is an important improvement 
which might help to establish top quark spin correlation at the LHC.

{\bf Acknowledgments:} I would like to thank A.~Brandenburg, A.~Ritz
and F.~Hubaut, P.~Pralavorio, E.~Monnier  from the Atlas collaboration for 
useful and encouraging discussions.  In particular I would like to
thank P.~Pralavorio for pointing out the significance of the invariant
mass cut.

%\bibliographystyle{NuclPhys+href}
%\bibliography{literatur}
\newcommand{\zp}{Z. Phys. }\def\as{\alpha_s }\newcommand{\prd}{Phys. Rev.
  }\newcommand{\pr}{Phys. Rev. }\newcommand{\prl}{Phys. Rev. Lett.
  }\newcommand{\npb}{Nucl. Phys. }\newcommand{\psnp}{Nucl. Phys. B (Proc.
  Suppl.) }\newcommand{\pl}{Phys. Lett. }\newcommand{\ap}{Ann. Phys.
  }\newcommand{\cmp}{Commun. Math. Phys. }\newcommand{\prep}{Phys. Rep.
  }\newcommand{\jmp}{J. Math. Phys. }\newcommand{\rmp}{Rev. Mod. Phys. }

\end{document}